\documentclass[journal,9pt]{IEEEtran}

\usepackage{amsmath,amsfonts}
\usepackage{algorithmic}
\usepackage{algorithm}
\usepackage{array}
\usepackage{subfigure}

\usepackage{textcomp}
\usepackage{stfloats}
\usepackage{url}
\usepackage{verbatim}
\usepackage{graphicx}
\usepackage{cite}
\usepackage[hidelinks]{hyperref}
\usepackage{xcolor}
\newcommand{\Fig}[1]{Fig.\,{\ref{#1}}}

\newcommand{\jj}{\mathrm{j}}

\begin{document}

\onecolumn
\newpage
\thispagestyle{empty}

\textbf{Copyright:}
\copyright 2025 IEEE.  Personal use of this material is permitted.  Permission from IEEE must be obtained for all other uses, in any current or future media, including reprinting/republishing this material for advertising or promotional purposes, creating new collective works, for resale or redistribution to servers or lists, or reuse of any copyrighted component of this work in other works.\\

\textbf{Disclaimer:} This work has been published in \textit{IEEE Transactions on Antennas and Propagation}. \\

Citation information: DOI 10.1109/TAP.2025.3602115

\newpage

\twocolumn

\setcounter{page}{1}

\title{\huge{Transfer ABCD Matrix  for Time-Varying Media and Time Crystals}}

\author{\Large{Carlos Molero, Pablo. H. Zapata-Cano, and Antonio Alex-Amor}

\thanks{C. Molero is with the Department of Electronic and Electromagnetism, Faculty of Physics, University of Seville, 41012, Seville, Spain (email: cmolero1@us.es)\\
P. H. Zapata-Cano is the Department of Signal Theory, Telematics and Communications, Research Centre for Information and 
Communication Technologies (CITIC-UGR), Universidad de Granada, Granada, Spain. (email:  pablozapata@ugr.es).\\ Antonio Alex-Amor is with the Department of Electronic and Communication Technology, RFCAS Research Group,  Universidad Autónoma de Madrid, 28049 Madrid, Spain (email: antonio.alex@uam.es).}}



\maketitle

\begin{abstract}
This paper introduces a formal description of the transfer ABCD parameters in time-varying electromagnetic systems. The formal description comes after the rearrangement of the electric displacement field $D$ and magnetic flux density field $B$ at the inputs and outputs of the temporal system based on the time-varying boundary conditions. Then, we derive the ABCD parameters of a temporal transmission line, i.e., a temporal slab, and compute the associated scattering parameters (reflection and transmission coefficients). The results presented here open up an alternative way, based on network theory, to analyze multilayer temporal configurations. Moreover, we show that the ABCD parameters can be used to compute the dispersion diagram ($\omega$ vs $k$) of time crystals.     

\end{abstract}

\begin{IEEEkeywords}
Network analysis, time-varying systems, ABCD parameters, scattering parameters, S-matrix, temporal slabs, time crystal.
\end{IEEEkeywords}

\section{Introduction}
\IEEEPARstart{M}{axwell's} equations describe with precision any electric and magnetic phenomena, or the interaction between both, involving classical field theory \cite{EnghetaMaxwell15}. This includes the understanding and control of electromagnetic (EM) wave propagation, radiation, wave guiding, light-matter interactions at the human and nano scales, the production of basic electronic components such as resistors, inductors, capacitors or transformers, to name a few, the use of metamaterials,  and many other interesting applications \cite{Oliner1984, PhotonicsBook, Balanis, CapolinoMetamaterials, Maci2024}. Certainly, Maxwell's equations have transcended and transformed our society.

In a given electromagnetic problem, Maxwell's equations provide a complete solution of it; that is, the electric and magnetic fields can be determined at all points in space and time. Nonetheless, the extraction of fields in complex scenarios involving intricate geometries, multi-scale elements, and/or more exotic materials such as graphene may suppose a challenging, even prohibitive, task in many occasions. In this context, the recent popularization of space-time photonics, with the introduction of time-varying materials, has added additional complexity to the analysis with respect to classical time-invariant and time-harmonic systems  \cite{caloz2019spacetime, caloz2019spacetimeII, galiffi2022photonics}.   

It is in this context where \emph{network analysis} or network theory emerges as an alternative to reduce computational burden and mathematical complexity. Network analysis seeks to analyze general EM problems in terms of distributed elements (transmission lines) and lumped elements via circuit theory: distributed elements generally model wave propagation, while lumped elements do so with junctions and discontinuities between media. Historically, the microwave and antenna communities have benefited from its use in the analysis and design of widely-used radiofrequency devices such as filters, waveguides,  amplifiers, couplers, and circulators \cite{pozar}. Similarly, the photonics and optics communities also make regular use of this tool, for example, to perform ray tracing computations under paraxial assumptions   \cite{Kogelnik66, Fienup24}.

In network analysis, a microwave or photonic network can be defined as a $N$-port system, whose inputs and outputs are interconnected via a governing matrix \cite{pozar}. Impedance $[Z]$ (admittance $[Y]$) matrices are widely used solutions to relate the input voltages (currents) and output currents (voltages) in the network. Similarly, the scattering matrix $[S]$ does so with incident and reflected power waves. However, in many practical scenarios, the number of ports is reduced to $N=2$ (single input and single output). In this context, the use of ABCD parameters becomes particularly relevant. 

The internal configuration of the ABCD matrix allows us to cascade multiple stages in the system by simply multiplying all individual ABCD matrices, a fact that the $[Z]$, $[Y]$, and $[S]$ matrices do not allow. For this reason, ABCD parameters are quite popular in  microwave engineering and have been intensively used for the production of 
filters \cite{Chen2010}, transistor amplifiers \cite{Thorson1999}, design of multilayer frequency selective surfaces (FSS) without and with cross-polarization terms \cite{alex2021exploring, molero2021cross}, or to simply compute the dispersion properties in periodic structures \cite{Mesa2021}. 

From a microwave theory perspective, the ABCD-matrix system can be seen as a ``parameter black box" that links the input and output of a certain temporal system. In fact, treating a complex system as a black box is a very powerful tool that has been used historically in the context of engineering and, more specifically, in the electronics, microwave and antenna communities. This has been the case for the analysis and design of conventional time-invariant microwave components such as filters, amplifiers, power dividers, mixers, transmission lines, or general waveguiding devices \cite{pozar}. Furthermore, the use of time modulation can be exploited in many engineering contexts, providing extra degrees of freedom and enhanced functionalities for applications such as MIMO communications or multi-target radar detection \cite{Li2025tap}.

Despite the well-known foundations of ABCD parameters and their vast application in time-invariant engineering scenarios, the formalization of an analogous framework suited to \emph{time-varying} EM problems is an open and evolving area of study \cite{EnghetaGaldi2021, Cummer2019, XuMaiWerner2022}.  In this work,  we exploit microwave theory concepts and the temporal boundary conditions of the electric displacement field $\mathbf{D}$ and magnetic flux density field $\mathbf{B}$ to rearrange the  input and output of the two-port time-varying system in a way that the resulting transfer matrix can be cascaded in a similar manner to conventional ABCD parameters. Then, we extract the ABCD parameters of a temporal transmission line (a temporal slab), and connect them with the associated scattering parameters and the computation of the dispersion diagram ($\omega$ vs $k$) in discrete and continuous time crystals. This formalism offers advantages compared to other matrix-based methodologies reported in the literature for time varying slabs \cite{Ramaccia2021, Milford2020, Leger2019}, specially due to the substantial reduction of the number of matrices needed to compute the electromagnetic response of a time multi-layer system. Furthermore, the formalism is useful to provide valuable and specific physical insight to understand energetic and momentum processes in time-varying slabs.

\section{Temporal Transfer ABCD Matrix}

\begin{figure*}[!t]
	\centering
\subfigure{\includegraphics[width= 1.70\columnwidth]{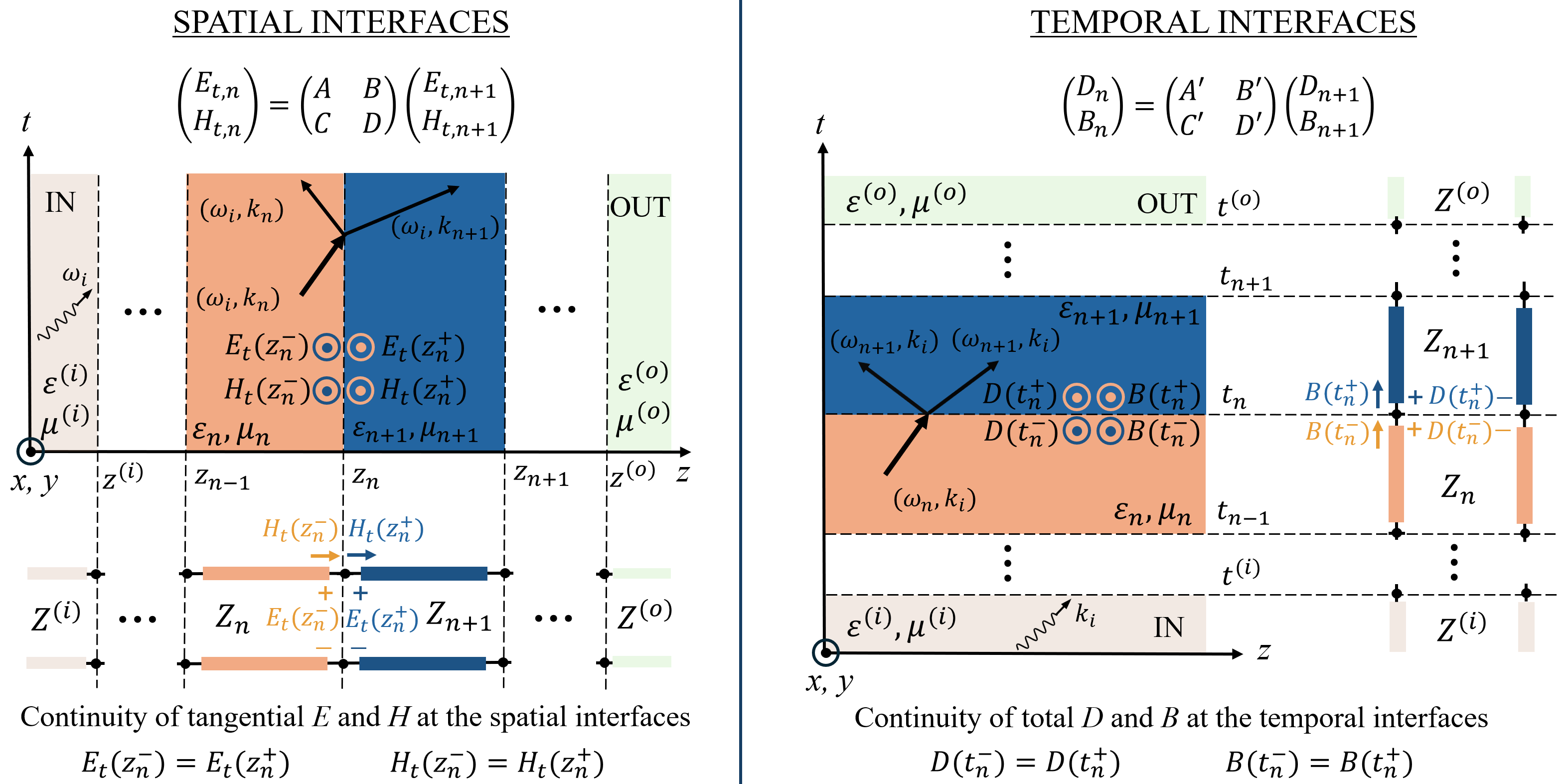}}
	\caption{Use of transfer ABCD parameters in multilayer spatial and temporal interfaces. In both scenarios, ABCD matrices can be used to connect the EM fields at the input and output of the system. By exploiting the continuity of tangential \{$E_t$,$H_t$\} (total  \{$D$, $B$\}) fields in the spatial (temporal) problems, the fields can be rearranged such that the ABCD matrix of each layer can be multiplied to obtain the global response of the time-invariant (time-varying) system. The multilayer composite is formed by $N+2$ layers: the input layer [$(i)$, $n=0$], $N$ intermediate layers, and the output layer [$(o)$, $n=N+1$].    }
	\label{fig1}
\end{figure*}

The starting point to derive the ABCD matrix comes from the identification of an analogy between purely spatial and purely temporal electromagnetic problems for multilayer composites made of linear, homogeneous, isotropic, and non-dispersive materials. A sketch of the spatial and temporal scenario is shown in Fig.\ref{fig1}. 

In a purely spatial scenario (left panel in Fig.~\ref{fig1}), a multilayer composite is defined as a cascade of individual layers with different permittivity/permeability $\varepsilon_{n}, \mu_{n}$ and thickness $d_{n}$. When a plane wave impinges on it, the boundary conditions at the dielectric interfaces impose the continuity of the tangential electric- and magnetic-field vectors. Under spatially-varying boundary conditions, the frequency of the system  keeps invariant, and identical to the impinging frequency $\omega_i$, while the wavevector changes according to $k_{n} = \omega_i \sqrt{\varepsilon_{ n}\mu_{n}}$. The invariance of the frequency leads us to tackle the problem in the temporal steady state: the time evolution becomes harmonic, $\text{e}^{\text{j}\omega t}$, and can be omitted in the calculations. 

As mentioned above, the ABCD-matrix formalism has proven to be quite efficient in characterizing cascade connections. As it is well known \cite{pozar}, the propagation of waves through a spatial dielectric layer of thickness $d_n$ admits to be represented by the following matrix:
\begin{equation} \label{spatialABCD}
\begin{pmatrix}
E_{n - 1}  \\
H_{n - 1} 
\end{pmatrix} = 
    \begin{pmatrix}
\cos(k_{n}d_{n}) & \text{j}Z_{n} \sin (k_{n}d_{n}) \\
\text{j} 1/Z_{n} \sin(k_{n}d_{n}) & \cos(k_{n}d_{n}) 
\end{pmatrix} 
\begin{pmatrix}
E_{n}  \\
H_{n} 
\end{pmatrix}
\end{equation}
where $E_{n-1}, H_{n-1}, E_{n}, H_{n}$ denote the electric and magnetic fields in the $n$-th order interface, with $Z_{n} = E_{n}/H_{n}$ being the wave impedance in medium $n$. The great advantage of the ABCD formalism with respect to others comes from the fact that the global matrix of the whole multilayer composite is the result of the product of the individual ABCD matrices. Thus, computation and analysis are simple and intuitive. Notice that \eqref{spatialABCD} coincides with the ABCD matrix of a conventional transmission line \cite{pozar}. This is why the resolution of multilayer systems is formally equivalent to solving a problem involving cascade of transmission lines. In these cases, the notation is expressed via voltages/currents instead of electric/magnetic fields. 

In a time-varying multilayer composite (right panel in Fig.~\ref{fig1}), the medium is uniform in space and progressively changes its properties in time at instants $t_n$.  At the time interfaces, the temporal boundary conditions enforce that the electric and magnetic flux vectors $D, B$ are continuous. In contrast to the spatial case, the wavevector $k$ is invariant under time-varying conditions, whereas the frequency shifts from one medium to the other according to $\omega_{n} = k_i /\sqrt{\varepsilon_{n} \mu_{n}}$ ($k_i$ is the wavevector in the first layer).
To address the problem with the ABCD formalism, it is necessary first to define a \emph{spatial steady state}, reached when the system varies harmonically in space as $\textrm{e}^{-\text{j}k z}$. As in the spatially-varying case, the $\textrm{e}^{-\text{j}k z}$ dependence here can be omitted in the time-varying scenario to simplify calculations. 

The apparent duality between both cases leads to the definition of an analogous ABCD-matrix formalism for time-varying multilayer composites. This matrix would look as 
\begin{equation} \label{generalABCD}
\begin{pmatrix}
D_{n - 1}  \\
B_{n - 1} 
\end{pmatrix} = 
    \begin{pmatrix}
A' & B' \\
C' & D' 
\end{pmatrix}
\begin{pmatrix}
D_{n}  \\
B_{n} 
\end{pmatrix}\,,
\end{equation} 
where $D_{n-1}, B_{n-1}, D_{n}, B_{n}$ are the electric/magnetic flux vectors in the $n$th-order interface. In the following, we use the prime notation ($'$) to identify the ABCD parameters and avoid confusion with vectors $D$ and $B$. As we will see below, the result of the matrix allows us to indentify a temporal transmission line.  \\

\subsection{Temporal Transmission Line (Temporal Slab)}

Let us consider the scenario depicted in the right panel of Fig. \ref{fig1}. In the scenario, a certain medium or material changes its electrical (or magnetic) properties over time at $t=t_{n-1}$. After a short period, the medium changes again its properties at $t=t_n$. The temporal region between $t_{n-1}$ and $t_n$ defines a \emph{temporal slab}, a homogeneous medium that covers the whole space and is present for a limited time. Analogously to a spatial slab, a temporal slab is defined by its temporal length $T = t_n - t_{n-1}$, as well as its permittivity $\varepsilon_{n}$, permeability $\mu_{n}$ and the associated wave impedance $Z_{n}$. Please note that the wave impedance is defined as $Z_n = B_n/D_n$ in the $n$-th temporal slab. This definition results from a proper combination between the impedance definition $Z_{n} = E_{n}/H_{n} = \sqrt{\mu_{n} / \varepsilon_{n}}$, and the constitutive relations $D_{n} = \varepsilon_{n} E_{n}$, $B_{n} = \mu_{n} H_{n}$ .

A temporal slab can be formally described as a \emph{temporal} transmission line, in analogy to the conventional spatial transmission line \cite{pozar}. A temporal transmission line is an object that connects two time instants $t_1$ and $t_2$. For simplicity, we consider $t_1 = 0$ and $t_2 = T$. Without loss of generality, the $D$ and $B$ fields can be represented at the input of the temporal transmission line in terms of forward ($+$) and backward ($-$) waves,
\begin{subequations} \label{D1}
    \begin{equation} 
    D_{1} = D(t=0) = D^{+} + D^{-}\, , 
    \end{equation}
    \begin{equation} \label{B1}
    B_{1} = B(t=0) = B^{+} + B^{-}\, , 
    \end{equation}
\end{subequations}
and at the output  as
\begin{subequations} \label{Dn+1Bn+1}
    \begin{equation}
    D_{2} = D(t=T) = D^{+}\mathrm{e}^{\jj \omega T} + D^{-}\mathrm{e}^{-\jj\omega T}\, , 
    \end{equation}
    \begin{equation}
    B_{2} = B(t=T) = B^{+}\mathrm{e}^{\jj \omega T} + B^{-}\mathrm{e}^{-\jj\omega T} , 
    \end{equation}
\end{subequations}

Since we work with linear, isotropic and non-dispersive media, $D(t) = \varepsilon(t) E(t)$ and $B(t) = \mu(t) H(t)$ must hold. Thus, the coefficients $D^+$, $D^-$, $B^+$ and $B^-$ are related as follows
\begin{subequations} \label{D+D-}
    \begin{equation}
        D^+ = B^+/Z,
    \end{equation}
    \begin{equation}
        D^- = -B^-/Z,
    \end{equation}
\end{subequations}
with $Z = \sqrt{\mu / \varepsilon}$. The combination of eqs. \eqref{D1}, \eqref{Dn+1Bn+1} and \eqref{D+D-} leads to the following relations
\begin{subequations} \label{B+B-}
    \begin{equation} \label{B+}
        B^+ = \left(\frac{B_{2} + Z D_{2}}{2}\right) \mathrm{e}^{-\jj \omega T},
    \end{equation}
    \begin{equation} \label{B-}
        B^- = \left(\frac{B_{2} - Z D_{2}}{2}\right) \mathrm{e}^{\jj \omega T}.
    \end{equation}
\end{subequations}
 After replacing eq. \eqref{B+B-} into the expression for $D_1$ in \eqref{D1}, we realize that fields $D_1$, $D_{2}$ and  $B_{2}$ are related as 
\begin{equation}
    D_1 = \cos(\omega T) D_{2} - \jj \frac{1}{Z} \sin(\omega T) B_{2}
\end{equation}
Similarly, the replacement of eq. \eqref{B+B-} into the expression for $B_1$ leads to
\begin{equation}
    B_1 = -\jj Z \sin(\omega T) D_2 + \cos(\omega T) B_{2}
\end{equation}

The connection between the fields at the two ports, input (1) and output (2), of the temporal transmission line can be better rearranged into matrix form:
\begin{equation} \label{ABCD_line}
    \begin{pmatrix}
D_{1}  \\
B_{1} 
\end{pmatrix} = 
    \begin{pmatrix}
\cos(\omega T) & -\jj \frac{1}{Z} \sin(\omega T) \\
-\jj Z \sin(\omega T) & \cos(\omega T) 
\end{pmatrix}
\begin{pmatrix}
D_{2}  \\
B_{2}
\end{pmatrix} \, , 
\end{equation}
thus revealing that $A' = \cos(\omega T)$, $B' = -\jj \frac{1}{Z} \sin(\omega T)$,  $C' = -\jj Z \sin(\omega T)$, and $D' = \cos(\omega T)$ are the transfer ABCD parameters of the temporal transmission line (temporal slab). Note that the frequency $\omega$ in eq.~\eqref{ABCD_line} varies across temporal slabs, while $k$ remains invariant. In general, the frequency in the $n$-th slab is given by $\omega = k / \sqrt{\varepsilon_n \mu_n}$.

It can be rapidly checked that the network is reciprocal ($A'D' -B'C' = 1$) and symmetrical ($A' = D'$) in terms of electric and magnetic flux densities $D, B$. This \emph{temporal} or \emph{momentum} reciprocity does not necessarily conduct to the reciprocity for $E, H$ (spatial reciprocity), and vice versa. In addition, it is worth highlighting the units of the four ABCD parameters: $A'$ [dimensionless], $B'$ [$\Omega^{-1}$], $C'$ [$\Omega$], and $D'$ [dimensionless]. This represents a difference with other matrix formalisms that describe time-varying systems \cite{Ramaccia2021, Leger2019}.

As in the spatial case, the transfer parameters can be cascaded. Thus, the overall response  of a time-varying system formed by $N$ connected temporal transmission lines ($N$ temporal slabs), each of those with characteristic impedance $Z_n$, will be simply extracted after cascading (multiplying) the transmission matrices of each stage. 

In the ABCD format, the number of matrices to multiply coincides with the number of temporal slabs $N$. Notice that this entails significant differences with other matrix formalisms reported in the literature, such as those in \cite{Ramaccia2021, Leger2019}. There, the authors present a sound matrix approach that connects forward and backward  $E$ waves, in a similar fashion to using scattering parameters, instead of directly working with $D$ and $B$. Since $E$ is not continuous across time interfaces, additional matrices (called matching/interface matrices) must be added for the entire calculation. For a temporal multilayer made of $N$ slabs, these methods typically need to employ $2N + 1$ matrices, in contrast to the method here presented. 

Another interesting ABCD-matrix approach is  reported in \cite{Milford2020}, mainly conceived for time-periodic circuit elements. The authors work with ABCD matrices by manipulating voltages and currents in the network. Some connections are made with electromagnetics, although the authors predominantly describe circuits and do not examine electromagnetic fields in depth. Moreover, the computation of the solutions in their formalism requires the resolution of \emph{infinite-dimensional} determinants, whose convergence may result problematic. Finally, it is worth remarking that our method provides intuitive physical insight concerning energy and momentum considerations.   
\subsection{Associated Scattering Parameters}

The analogy between spatial and temporal problems gives rise to the possibility of extracting the S-parameters associated with the temporal ABCD matrix. Now, the formal definition of the scattering matrix (S-matrix) takes the $D^{\pm}$, forward ($+$) and backward ($-$), waves at the input (1) and output (2) media. This contrasts to the spatial case, based on electric-field counterparts $E^{\pm}$. In both cases, however, there is a direct correspondence between them thanks to the constitutive relation $D^{\pm} = \varepsilon E^{\pm}$. The temporal S-matrix can be defined as:
\begin{equation} \label{Smatrix}
    \begin{pmatrix}
D_{2}^{-}  \\
D_{2}^{+} 
\end{pmatrix} = 
    \begin{pmatrix}
S'_{11} & S'_{12} \\
S'_{21} & S'_{22}
\end{pmatrix}
\begin{pmatrix}
D_{1}^{+}  \\
D_{1}^{-}
\end{pmatrix} \, , 
\end{equation}
with $D_{1}^{\pm}$ being the original forward/backward waves before the first slab switching (or input media), and $D_{2}^{\pm}$ the final forward/backward waves after the last slab switching (output media). The matrix coefficients, after a few calculations, are expressed as:
\begin{align}
\label{S11}S'_{11} &= \triangle [-A' - B'Z_{0} + C'/Z_{0} + D']\\
\label{S12} S'_{12} &= \triangle [-A' + B'Z_{0} - C'/Z_{0} + D']\\
\label{S21} S'_{21} &= \triangle [A' - B'Z_{0} - C'/Z_{0} + D'] \\
\label{S22}S'_{22} &= \triangle [A' + B'Z_{0} + C'/Z_{0} + D']
\end{align}
with $\triangle = [2A'D' - 2B'C']^{-1}$ and $Z_0 = Z^{(i)} = Z^{(o)}$. We use the prime notation to identify the scattering parameters in the temporal problem and thus distinguish them from conventional time-invariant scenarios. 

In the above definition, the parameters $S'_{11}$ ($S'_{21}$) would represent the ratio of the backward (forward) $D$-waves at the output port (2) and an incident forward $D$-wave at the input port (1). Similar physical reasoning is used to define $S'_{12}$ and $S'_{22}$, but with respect to an incident backward wave at the input port (1).
   
By assuming identical input/output media, the corresponding  S-matrix for the electric field coincides with those in \eqref{S11}-\eqref{S22}, 
\begin{equation}
S'_{ij} = \frac{D_{2}^{\pm}}{D_{1}^{\pm}} = \frac{E_{2}^{\pm}}{E_{1}^{\pm}}\,.
\end{equation}
The definition of a more general framework of relationships involving different input/output media, $Z^{(i)} \ne Z^{(o)}$, is still under discussion and their physical implications are now focus of research.


\subsection{Considerations about Conservation of Momentum Density }

The instantaneous density momentum in temporal systems is the variable defined by $\mathbf{g}(t, z) =\mathbf{D}(t, z) \times \mathbf{B}(t, z)$, in complete analogy to the instantaneous Poynting vector for spatial systems $\mathbf{S}(t, z) = \mathbf{E}(t, z) \times \mathbf{H}(t, z)$. Although both quantities can formally be defined in both systems, $\mathbf{g}(t, z)$ and $\mathbf{S}(t, z)$ in temporal and spatial problems play an equivalent role, respectively.  

As is well known, the Poynting vector $\mathbf{S}(t,z)$ [W/m$^2$] indicates the power per unit area that the EM wave carries and points towards the  direction in which the wave propagates. In the spatial multilayer composite, which is a time-harmonic system, the time-average of $\mathbf{S}(t, z)$ is extracted as
\begin{equation}
\langle \mathbf{S}(z)\rangle_{t} = \frac{1}{2} \text{Re}\{\mathbf{E}(z)\times \mathbf{H}(z)^{*}\}\,.
\end{equation}
The terms $\mathbf{E}(z)$ and $\mathbf{H}(z)$ represent the complex-valued electric and magnetic fields in phasor form. The quantity $\langle \mathbf{S}(z)\rangle_{t}$ denotes the average power per unit area and per time period at each spatial position $z$. In passive and lossless systems, this result manifests in the form of the energy conservation across the layers, leading to the expression
\begin{equation}
E(z^{(i)})\cdot [H(z^{(i)})]^{*} = 
\underset{\text{Intermediate layers}}{\cdots} = 
E(z^{(o)})\cdot [H(z^{(o)})]^{*}\, ,
\end{equation}
with $i, o$ indicating the input/output layers of the  spatial multilayer composite. 

A similar rationale applies in the context of temporal multilayer structures for the instantaneous  momentum density vector $\mathbf{g}(t, z)$. However, some subtle differences must now be taken into account. The system varies harmonically in space, not in time, thus the \emph{space-average} momentum density is 
\begin{equation}
\langle \mathbf{g}(t)\rangle_{z} = \frac{1}{2}\text{Re}\{\mathbf{D}(t)\times \mathbf{B}(t)^{*}\}.
\end{equation}
The spatial average is physically interpreted, at any time instant $t$, as the summation of the momenta \emph{collected} in different spatial points along an interval of a wavelength $\lambda$. 
Analogously to the time-average Poynting vector $\langle \mathbf{S}(z)\rangle_{t}$, it can be checked that the space-average momentum density $\langle \mathbf{g}(t)\rangle_{z}$ is conserved in lossless temporal multilayer composites. Thus,
\begin{equation}
D(t^{(i)}) \cdot [B(t^{(i)})]^{*} = \underset{\text{Intermediate layers}}{\cdots} = D(t^{(o)}) \cdot [B(t^{(o)})]^{*}\,.
\end{equation}
 The above has direct implications on power and energy  in time-varying systems. In fact, it can be readily demonstrated that, if the input and output media are electrically/magnetically different ($\varepsilon^{(i)} \neq \varepsilon^{(o)}$ or $\mu^{(i)} \neq \mu^{(o)}$), the energy is not conserved between the input and output layers independently of the number of intermediate temporal layers and electrical properties of these. Likewise, it is expected that the energy transferred from the input to the output is conserved, independently of the phenomena occurred in the intermediate layers, if the input and output layers have the same electrical properties ($\varepsilon^{(i)} = \varepsilon^{(o)}$ and $\mu^{(i)} = \mu^{(o)}$).

\subsection{Time Crystal: Infinite  Temporal Transmission Lines}

Analogously to a spatial crystal, a time crystal is constituted by a periodic repetition of elements in time. Likewise, all information can be extracted from the unitary element that is repeated over time, the so-called \emph{unit cell}, and the application of the Floquet-Bloch theorem \cite{Floquet1883}. Probably, the simplest manner to implement a time crystal is simply cascading temporal transmission lines (temporal slabs) in a time-periodic scheme. In this scenario, we consider that all spatial points in the whole medium change their electrical properties simultaneously. 

Let us know analyze the  case described above: the interconnection of infinite identical time-varying systems in a time-periodic scheme. The temporal unit cell of period $T$ admits to be represented by a transfer ABCD matrix whose fields fulfill $D_{n+1} = D_{n}\mathrm{e}^{\jj \omega T}$ and $B_{n+1} = B_{n}\mathrm{e}^{\jj \omega T}$, according to the Bloch theorem. As a consequence, the general expression for a temporal transfer ABCD matrix given in eq. \eqref{generalABCD} can be rewritten as the following eigenvalue problem,
\begin{equation}
    \begin{pmatrix}
D_{n}  \\
B_{n} 
\end{pmatrix} 
\left[
\begin{pmatrix}
A' & B' \\
C' & D' 
\end{pmatrix} \mathrm{e}^{\jj \omega T} - \begin{pmatrix}
1 & 0 \\
0 & 1 
\end{pmatrix}
\right] 
 = 0\, ,
\end{equation}
that has a non-trivial solution when the determinant of the matrix nulls. This leads to the equation
\begin{equation}
    \mathrm{e}^{-\jj 2\omega T} - \mathrm{e}^{-\jj \omega T} (A' + D') + A'D' - B'C' = 0\, ,
\end{equation}
whose solution is
\begin{equation} \label{omega_dispersion}
    \omega = \frac{\jj \ln(a)}{T}\,
\end{equation}
with $a = 0.5 (\pm \sqrt{A'^2 - 2A'D' + 4B'C' + D'^2} + A' + D')$. Eq. \eqref{omega_dispersion} offers a pathway to compute the dispersion diagram, $\omega$ vs $k$, in time crystals. In the above, note that $A'D' - B'C' = 1$. The wavenumber dependence in eq. \eqref{omega_dispersion} is in the auxiliary parameter $a = a(k)$, since the ABCD parameters are dependent on the selected $k$ as well.

The advantage of the approach of the present paper is the possibility of extracting analytical expressions to compute $\omega(k)$ diagrams. Other methodologies in the literature, such as that reported in \cite{Halevi2016}, address the problem from a numerical perspective by using complex root-finding algorithms.  Our methodology provides formal analytical expressions for $\omega(k)$, including complex-valued frequency solutions. This is a significant advantage since evanescent- and, especially, growing-field solutions manifested via the imaginary part of the frequency can not be lightly disregarded in time-varying systems \cite{Silveirinha2014}. 
\section{Results}
The usage of transfer ABCD matrices is a simple, but really useful, strategy to analyze a wide variety of scenarios involving time-varying parameters and fields. In the following, we will show that the present methodology offers a complementary path to the existing tools for the analysis and design of time crystals and time-varying systems internally constituted by temporal slabs.

\subsection{Temporal multilayer slab}
\begin{figure}[!h]
	\centering
\subfigure{\includegraphics[width=1 \columnwidth]{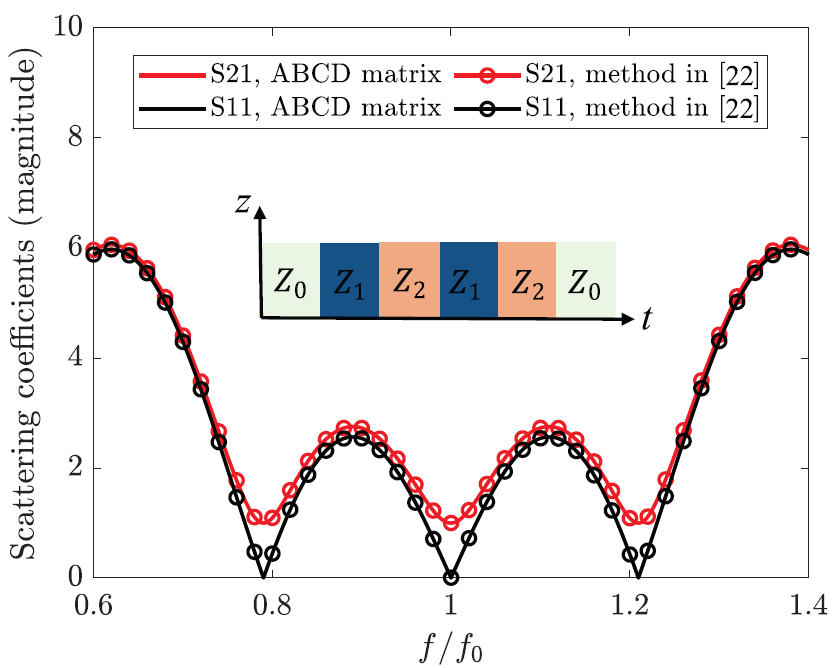}}
	\caption{Scattering parameters of an equally travel-distance multi-layered temporal structure with identical input and output media. $N=4$, $Z_1=Z_0/9$, $Z_2=Z_0/3$, and $Z^{(i)} = Z^{(o)} = Z_0$. Frequencies are normalized to the central frequency $f_0$.}
	\label{fig:multilayer}
\end{figure}

On a first instance, the proposed methodology is applied to the scattering analysis of multilayer temporal structures. In particular, the proposed scenario consist on the concatenation of several temporal dielectric ($\mu_r=1$) slabs of alternating impedances $Z_1$ and $Z_2$, being the input and output media air ($Z^{(i)}=Z^{(o)} = Z_0$). Regarding the modulation times, they are set according to the equally travel-distance multi-layered temporal structure paradigm introduced in \cite{Ramaccia2021}. Thus, they all share the same propagation distance $L_s = \lambda/2\, (\delta = \pi)$, where ${L_s = c Z T_n}$, being $T_n$ the temporal length of the $n$-th slab. Fig. \ref{fig:multilayer} depicts the $S_{11}$ and $S_{21}$ parameters calculated both with the hereby proposed methodology and the method presented in \cite{Ramaccia2021}.

The first notable conclusion one can extract from Fig.~\ref{fig:multilayer}  is that the results provided by the ABCD transfer matrix are identical to the ones given by the formulation in \cite{Ramaccia2021}, which validates the accuracy of the method in this scenario. As a direct consequence of the $L_s=\lambda/2$ condition, the phase delay induced by each temporal slab is $\pi$. Under these circumstances, the forward- and backward-scattered waves cancel out, making the whole structure transparent to the propagating wave. Moreover, having the same input and output impedances, added to the temporal periodicity of the structure, produces the vanishment of the scattered waves at a number of discrete frequency points ($\sim \{0.8,1,1.2\}f_{0}$ in the figure), which are directly correlated to the number of the temporal slabs of the system $N$. At these points, the reflection coefficient nulls, $|S_{11}| = 0$, and all the wave is transmitted, $|S_{21}|=1$, thus presenting the forward-scattered wave the same amplitude as the incident one while the backward wave disappears.

\subsection{Time Crystal}

\begin{figure}[!t]
\centering
\subfigure[]{
\includegraphics[width=0.9\columnwidth]{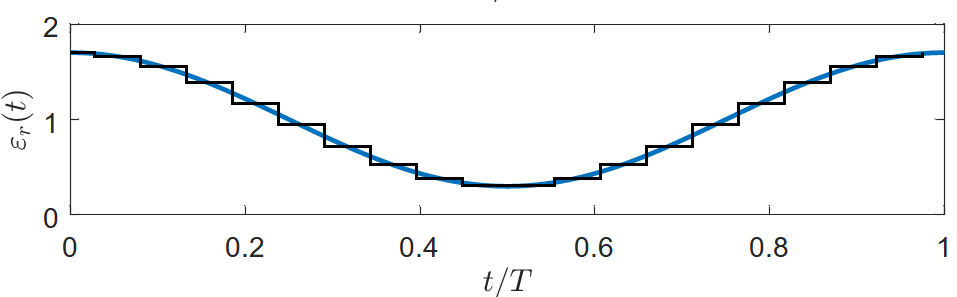}
}
\subfigure[]{
\includegraphics[width=0.95\columnwidth]{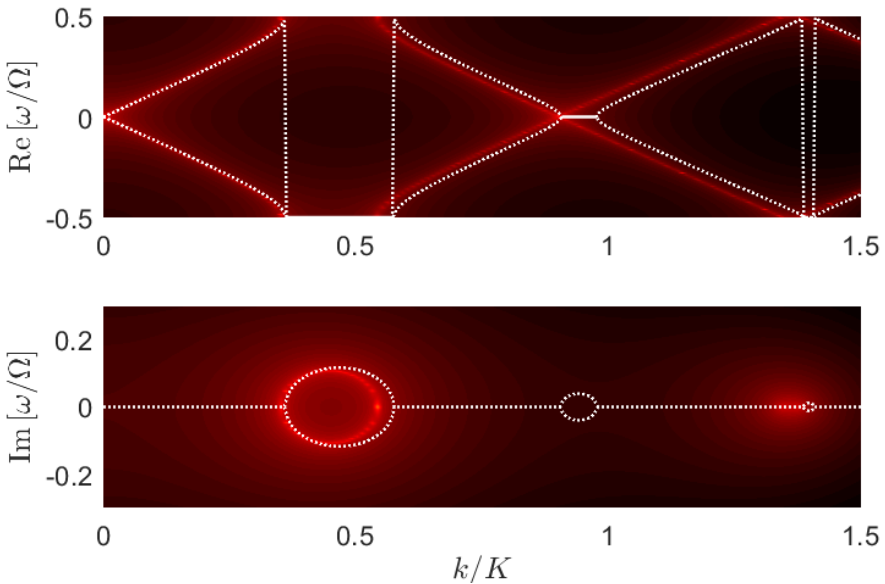}
}
\subfigure[]{
\includegraphics[width=0.95\columnwidth]{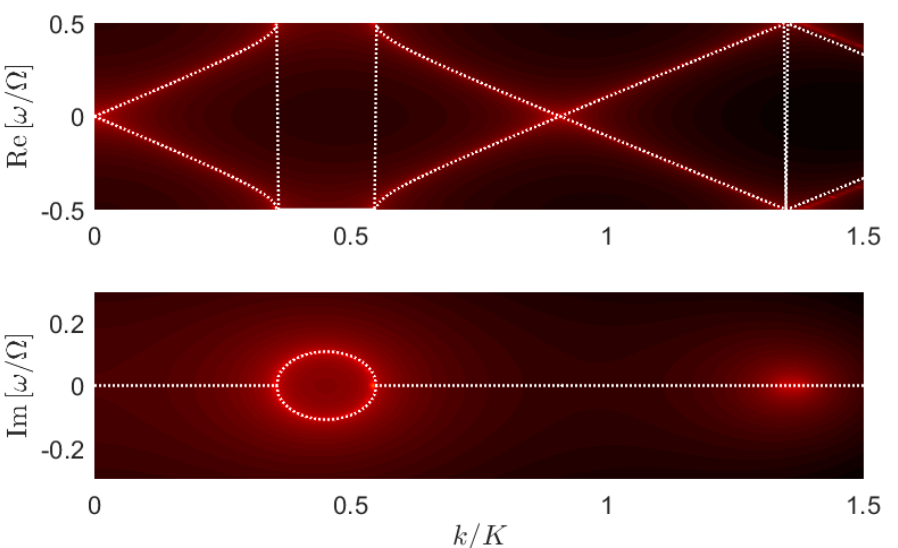}
}
\caption{Dispersion diagram, computed with the ABCD formalism (white dotted lines) and the exact numerical solution (colored plot), for a time crystal given by the temporal modulation $\varepsilon_r(t) = \varepsilon_{r0} \left[1 + m \cos(\Omega t) \right]$. (a) Discretization of the continuous temporal modulation (blue line) into $N$ steps (black lines). Dispersion diagram in the cases: (b) $N = 5$, (c) $N = 30$. Parameters: $\varepsilon_{r0} = 1$, $T = 0.1$ ns,  $m = 0.7$, and $\mu_r = 1$.  }
\label{TimeCrystal}
\end{figure}

In principle, the usage of transfer ABCD matrices via eq. \eqref{omega_dispersion} is originally intended for the analysis of discrete time crystals, i.e., for crystals formed by the interconnection of  transmission line sections. Nonetheless, when a continuous time crystal is discretized into $N$ small subsections, the electromagnetic response of both discretized and  continous time crystals should resemble as $N$ increases.

Let us consider a continous time crystal given by the time-periodic modulation $\varepsilon_r(t) = \varepsilon_{r0} \left[1 + m \cos(\Omega t) \right]$, with 
$ \varepsilon_{r0}$ being the DC component of the time modulation, 
$\Omega = 2\pi/T$ ($T$ is the temporal period of the time crystal), and $m$ is the modulation factor. For simplicity, we only consider a modulation of $\varepsilon_r(t)$ ($\mu_r = 1$), although a temporal magnetic response via $\mu_r(t)$ can be straightforwardly added. This type of time crystals has been successfully studied by Peter Halevi et al. \cite{zurita2009reflection, Halevi2016, Halevi2015}.  By taking advantage of the periodic nature of $\varepsilon_r(t)$ and $\mu_r(t)$, the authors expand the electrical parameters into complex Fourier series and solve the problem in the frequency domain. Alternatively, we can solve the problem by discretizing $\varepsilon_r(t)$ into $N$ small pieces and the subsequent use of eq. $\eqref{omega_dispersion}$ to compute the dispersion diagram. Thus, each piece represents a short temporal transmission line of length $T/N$ and impedance \linebreak $Z_n = \sqrt{\mu_r(nT)/\varepsilon_r(nT)}$, which is modeled via the transfer ABCD matrix in eq.  \eqref{ABCD_line}.

Fig.~\ref{TimeCrystal} exemplifies the analysis of a continuous time crystal of this kind, defined by the electrical parameters
$\mu_r = 1$ and $\varepsilon_r(t) =  1 + 0.7 \cos(2\pi 10^{10} t) $. Fig.~\ref{TimeCrystal}(a) shows the discretization of the photonic crystal into $N$ sections. Figs.~\ref{TimeCrystal}(b)-(c) show the normalized dispersion diagram (top panels for real frequencies and bottom panels for imaginary frequencies) in the cases of discretizing the time crystal into $N=5$ and $N=30$ sections, respectively. The normalizing factors are $\Omega$, previously defined, and $K = kc/(\Omega \sqrt{\varepsilon_{r0} \mu_r})$. In the figures, white dotted lines represent the solution extracted with the ABCD formalism and the colored plot (reddish lines with dark background) the exact numerical solution computed with Halevi's method \cite{Halevi2016}. As can be checked from Figs.~\ref{TimeCrystal}(b)-(c), the solution given by the ABCD formalism approaches the exact numerical solution as $N$ increases. For $N = 30$ steps, accurate solutions are obtained.

Finally, it is important to remark that the present method allows to compute both the real and imaginary parts of the frequency. In time crystals, imaginary frequency components refer to time decay/grow of the corresponding fields. By following the engineering notation ($\mathrm{e}^{+j\omega t}\, \mathrm{e}^{-jk z}$), positive imaginary frequencies in the dispersion diagram imply attenuation over time, while negative imaginary frequencies imply amplification. These solutions are typical from the bandgap regions, manifesting from $k/K \in [0.4, 0.55]$ in \Fig{TimeCrystal}(b)-(c). Moreover, it is expected that the present ABCD approach can be extended to the analysis of uniform-velocity space-time crystals, in a similar fashion to was carried out in \cite{CalozOrthogonal}.

\section{Conclusion}
In this paper, we explore a formal definition of the transfer ABCD parameters applied to two-port time-varying systems and time crystals. The definition comes from microwave theory concepts and the rearrangement of the fields $\mathbf{D}$ and $\mathbf{B}$ at the input and output, so that the temporal boundary conditions are fulfilled. Thus, the governing temporal transfer matrix [eq. \eqref{generalABCD}] can be cascaded in a similar fashion to conventional ABCD parameters. Then, we have computed the ABCD parameters of a temporal transmission line (temporal slab) in eq. \eqref{ABCD_line}, and its associated scattering parameters in eqs. \eqref{S11}-\eqref{S22}. Moreover, we have discussed on the continuity of the momentum density vector $\mathbf{D} \times \mathbf{B}$, and derived an analytical expression [eq. \eqref{omega_dispersion}] to compute the dispersion diagram ($\omega$ vs $k$) in time crystals via the ABCD formalism. Finally, we have tested the theory by computing the dispersion diagram of a time crystal and the scattering parameters of a temporal multilayer structure. The results presented in Section III validate the use of the proposed temporal transfer ABCD matrix and open up new possibilities in the analysis and design of time-varying systems.

\section*{Acknowledgments}

This work was supported by grant PID2020-112545RB-C54
funded by MCIN/AEI/10.13039/501100011033 and by European
Union NextGenerationEU/PRTR. It has also been supported
by TED2021-131699B-I00, PDC2023-145862-I00, and IJC2020-
043599-I, funded by MCIN/AEI/10.13039/501100011033 and by The
European Union NextGenerationEU/PRTR. It has also been supported in part by Consejería de Universidad, Investigación e Innovación of Junta de Andalucía through grant EMERGIA 23-00235".

\bibliographystyle{IEEEtran}

\bibliography{IEEEabrv, ref}


\vfill

\end{document}